%% file: Grains.tex
\magnification=\magstep1 \input macros
\def\normalbase{\baselineskip 12pt} \standardpage

\footline={\ifnum\pageno=1 \hfill \else\hss\tenrm\folio\hss\fi}
\tenstart
\centerline{\romtwelve Coarse grains: the emergence of space and order}

\bigskip
\centerline {L. S. Schulman}
\medskip
\centerline {\romnine Physics Department, Clarkson University}
\centerline {\romnine Potsdam, NY 13699-5820 USA}
\centerline{\romnine schulman@clarkson.edu}
\bigskip
\centerline {Bernard Gaveau}
 \medskip
\centerline{\romnine G\'eom\'etrie des \'equations aux  d\'eriv\'ees partielles et physique math\'ematique}
 \centerline{\romnine Universit\'e Paris 6, UFR 920, Case courrier 172, Tour 46, 5\`eme \'etage}
\centerline{\romnine 4, Place Jussieu, 75252 Paris C\'edex 05, FRANCE}
\centerline{\romnine gaveau@ccr.jussieu.fr}

\bigskip

{\narrower \romnine \baselineskip 10pt \def\it{\italnine} \def\bf{\boldnine}
\centerline{ABSTRACT} \vskip 2pt
The emergence of macroscopic variables can be effected through {\it coarse graining}. Despite practical and fundamental benefits conveyed by this partitioning of state space, the apparently subjective nature of the selection of coarse grains has been considered problematic. We provide objective selection methods, deriving from the existence of relatively slow dynamical time scales. Using a framework for nonequilibrium statistical mechanics developed by us, we show the emergence of both spatial variables and order parameters. Although significant objective criteria are introduced in the coarse graining, we do not provide a unique prescription. Most significantly, the grains, and by implication entropy, are only defined modulo a characteristic time scale of observation.
\par}
\header{1.~ Introduction}

Coarse grains lie at the foundations of statistical physics. They provide a technical cure for the famous paradoxes of the field, leading to mathematically rigorous justifications of irreversibility [\nr{\kac}]. And they make intuitive physical sense. A related issue is the profound distinction between work and heat, forms of energy that enter macroscopic and microscopic degrees of freedom, respectively. But this too is the issue of coarse graining, since one can look on macroscopic degrees of freedom as the labels of coarse grains.

Nevertheless, there is a severe unresolved problem. There is no law of Nature that defines the coarse grains. But there is a law of thermodynamics---the second---that presupposes them, in particular with respect to the heat/work distinction. The trouble with coarse grains though is that they seem subjective and technology dependent. One observer will smear over $10^{-6}\,$m and another, with better equipment, $10^{-8}\,$m. One observer will throw all Carbon into the same grain, another will distinguish $^{12}$C from $^{14}$\hbox{C}. The questions that arise from this are: 1) what picks the coarse grains, and 2) if there is arbitrariness, does it have physical implications, e.g., with respect to the second law? [\nr\mackey]

In this article we show how an objectively defined dynamics picks coarse grains. Mostly these grains turn out to be defined with respect to coordinate space, a result that many people who do coarse graining, whether for quantum or classical systems, take as natural (for example, see [\nr{\zeh}] or [\nr\hartle]). But as will be demonstrated, by looking at the dynamics we will find that space is not the only criterion for coarse graining. Under appropriate circumstances, one also gets an order parameter, magnetization in the example we consider. This order parameter is recognized as an {\it emergent\/} quantity, and it is gratifying that indeed it emerges as a macroscopic degree of freedom from the many microscopic variables. Although we will work entirely in a classical context, the results parallel those seen in quantum mechanics, where grains should also satisfy a decoherence condition. Thus Halliwell [\nr{\halliwell}] finds that local energy conservation makes local energy density a decohered quantity.

A related observation is that at many levels, from hydrodynamics to electrodynamics, one averages over microscopic variables to emerge with quantities such as density or polarization. By establishing macroscopic laws for these quantities (e.g., the Navier-Stokes equation), one confirms that these are indeed appropriate quantities. Our question in this paper goes beyond this process. It asks, if we did not know from extensive experience what variables to choose, what would pick them?

Our answer is basically, dynamics. The key to defining coarse grains is the existence of varying {\it time scales}. States are thrown into the same coarse grain if they are close to each other {\it dynamically}, meaning the system moves from one state to the other in a relatively short time. Note that in general our states will live in a large space, not necessarily related to ordinary configuration space (they are to be thought of as the microscopic states of a large system). It will turn out, however, that position does emerge from the coarse graining, essentially because particle-particle interactions are often local in space. We will argue nevertheless that temporal considerations are the essential criteria, not spatial. Part of this argument involves recalling the relation between coarse grains and irreversibility, for example following the exposition in Appendix A of [\nr{\effect}] (cf.\ \S7 below).

\aq{\master}\aq{\framework}\aq{\firstor}\aq{\signat}

We emphasize that the physics behind our idea is not new although it hardly could be said to enjoy universal consensus, nor has it been implemented, to our knowledge, in any systematic way. Our main goal is to show that the idea {\it can} be implemented within the context of our general method [\master--\signat] for treating nonequilibrium statistical mechanics. Here is what Landau and Lifshitz said about macroscopic variables [\nr\landau]:

\begingroup
\narrower\smallskip\NI
\ellip\ Owing to the comparative slowness of chemical reactions, equilibrium as regards the motion of the molecules will be reached \ellip\ more rapidly than equilibrium as regards \ellip\ the composition of the mixture. This enables us to regard the partial equilibria \ellip\ as equilibria at a given \ellip\ chemical composition.
              
The existence of partial equilibria leads to the concept of macroscopic states of a system. Whereas a mechanical microscopic description \ellip\
specifies the coordinates and momenta of every particle, \ellip\
a macroscopic description is one which specifies the mean values of the physical quantities determining a particular partial equilibrium, \ellip
\smallskip
\endgroup

As we will demonstrate, our nonequilibrium formalism provides a framework within which to translate this powerful physical idea into a specific coarse graining, hence a specific concept of ``macroscopic."

In Sec.\ 2 we introduce notation. Following that we develop notions of distance that arise from the dynamics alone. In Sections 4, 5 and 6 we give examples that examine these distances in a number of systems: Brownian motion (\S4), the Ising model (\S5) and heat flow (\S6). In Sec.\ 7 we discuss these results.

\header{2.~ Notation and framework}

As developed in [\master--\signat], we consider a system whose microscopic states take values in a state space $\Omega$ of cardinality $N<\infty$. The system's motion is a Markov process, $X(t)$, in $\Omega$ with transition matrix $R$. (N.B., some authors define this matrix as the transpose of our $R$.) Thus
$$ R_{xy}=``\Pr\bigl(x\leftarrow y)"=\Pr(X(t+1)=x\mid X(t)=y\bigr)\;. $$
We assume $R$ to be irreducible, so that it has a unique eigenvalue 1 with a strictly positive eigenvector $\pzero$:
$$ \sum_{x\in \Omega} R_{xy} = 1  \hbox{~~~~and~~~}
   \sum_{y\in \Omega} R_{xy} \pzero(y) = \pzero(x)>0 \quad \forall x \;. $$
The eigenvalues (\mquote{\lambda}) of $R$ are ordered by decreasing modulus (and increasing phase when applicable). Thus $\lambda_0\equiv 1 \geq |\lambda_1| \ge |\lambda_2| \ge \ldots$. The corresponding right and left eigenvectors are respectively $p_k$ and $A_k$, and satisfy
\numeq{\eigvals}{R p_k = \lambda_k p_k\;, \qquad A_k R = \lambda_k A_k\;,\quad k=0,1,\ldots \;. }%
Although $R$ may not be diagonalizable (and may require a Jordan form), we will assume that for the eigenvalues that concern us (those near 1) each eigenvalue possesses one or more eigenvectors. They are taken to be real. The orthonormality, $\langle A_k | p_\ell \rangle = \delta_{k \ell}$, still leaves a single multiplicative factor for each pair $(A_k,p_k)$. The stationary state $p_0$ is naturally normalized by $\sum p_0(x) = 1$, which fixes $A_0(x) = 1$ for all $x$. When detailed balance holds (which we do not assume in general), left and right eigenvectors are related by
$$  A_k(x)=p_k(x)/p_0(x) \;. $$
Consistent with this, we demand, for any $R$, that the $A$s be normalized by
$$ \sum_x p_0(x)\left(A_k(x)\right)^2 =1 \;.$$

\subheader{2.1 Coarse graining formalities}

Coarse graining is accomplished by lumping microstates into a macrostate. The coarse grains are (disjoint) subsets of $\Omega$ (covering all of $\Omega$), labeled $\xtilde$ and taking values in a space $\omtilde$. Each $x\in\Omega$ can be written $x=(\xtilde,u)$, with $u$ a {\it microscopic} label, identifying the particular microstate of $\xtilde$ that $x$ is. 
The coarse grained stationary distribution is $ \ptilde(\xtilde) = \sum_{u\in\xtilde} \pzero(\xtilde,u)$. The coarse grained transition matrix, $\rtilde$, governs transitions in $\omtilde$ and is given by
$$\rtilde_{\xtilde\ytilde}\equiv
  \sum_{u_1,u_2}R_{(\xtilde,u_2)(\ytilde,u_1)}\frac{\pzero(\ytilde,u_1)}{\ptilde(\ytilde)}$$
for $\xtilde\neq\ytilde$; each $\rtilde(x,x)$ is adjusted so that columns sum to one. With this definition, $\ptilde$ is the eigenvector of $\rtilde$ with eigenvalue one, currents add [\nr{\currentdef}] and other good properties are maintained. One can also rescale time.

In this paper we will not be concerned with the coarse graining itself, concentrating instead on how, in principle, one selects the coarse grains.

\headerandsubheader{3. Distance functions on $\Omega$}
                   {3.1 Left eigenvectors and dynamical proximity}

In [\firstor] we found a basic identity that allowed us to use eigenvalue near-degeneracy to arrive at near-reducibility of the transition matrix, interpreted as the breaking of a system into two or more phases. Here we use essentially the same identity in a more general way, with ``near-degeneracy" replaced by the existence of a sequence of time-scales reflected in a clustering of eigenvalues near 1 [\nr\noeigs].

For $t > 0$ an integer, we call $R^t$ the $t\ordinalth$ power of $R$. From \eq{\eigvals}, $\lambda_k^t A_k(y) =\sum_x A_k(x)(R^t)_{xy}$. The quantity $(R^t)_{xy}$ is the distribution function for a system that at time zero was definitely in the state $y$. Thus we write $p_y(x,t) \equiv(R^t)_{xy}$. 

Suppose that $\lambda_1,\lambda_2,\ldots,\lambda_p$ are close to 1 in magnitude. In contrast to [\firstor] we do not require subsequent eigenvalues to be significantly smaller. Consider the $N$ $p$-vectors $V(x) \equiv (A_1(x),\allowbreak A_2(x),\allowbreak \ldots, \allowbreak A_p(x))$ in \Rp. Call the set of these vectors $\cal A$ and their convex envelope \hullA. Since $\sum p_0(x)A_k(x)=0$ for all $k$, the $p$-vector 0 is in \hullA. Note that \hullA\ is of codimension 0 in \Rp\ [\nr\proveindep]. In [\firstor] we showed that there exist $p+1$ points, $z_1,z_2,\ldots,z_{p+1}$, such that $(A_1(z_\ell), A_2(z_\ell), \ldots, A_p(z_\ell))$ are the extremal points of the convex envelope (which contains 0) of all $N$ $p$-vectors. Call these extremal vectors $V_\ell \equiv V(z_\ell)$, $\ell=1, \ldots, p+1$. From the eigenvector property of the $A$s, it follows that for any $V(x)$, $x\in\Omega$
\numeq{\multiplelefteig}{(\lambda_1^t A_1(x), \lambda_2^t A_2(x), \ldots, \lambda_p^t A_p(x))
  =\sum_{y\in\Omega} p_x(y,t)(A_1(y), A_2(y), \ldots, A_p(y))  }%
We write the left hand side of \eq{\multiplelefteig} as $V(x)\Lambda^t $, with $\Lambda$ the diagonal ($p\negonemu\times p$) matrix of eigenvalues. Recalling that $\sum_y p_x (y,t)=1$, it is immediate that
\numeq{\VellminusV}{ V_\ell (1-\Lambda^t) =\sum_y p_{z_\ell}(y,t)\left ( V_\ell -V(y) \right) }%
In [\firstor] we showed that there exist linear forms $h_{\ell,j}$, with coefficients $\leq1$, such that
\numeq{\linearforms}{ h_{\ell,j}\left(V_\ell-V(y) \right) \geq 0 
\hbox{~~for~} \ell=1,\ldots,(p+1)
\hbox{~and~} j=1,\ldots,p \;,  }%
with equality only if $y=z_\ell$. \eq{\linearforms}, coupled with an assumption on $\lambda_{p+1}$, was the basis for our conclusions on the existence of $p+1$ phases. 

Our present use of \eq{\linearforms} is to argue for the near equality of the values of the left eigenvectors, $A$, when points are dynamically close. This will motivate our definition of a distance function on $\Omega$. Consider the left hand side of \eq\VellminusV. Because $\lambda_1, \dots, \lambda_p$ are by  assumption close to 1, this will be small. This requires the right hand side to be small. For given $\ell$ we apply the $p$ linear forms $h_{\ell,j}$, $j=1,\ldots, p$ to the equation. The sum now involves only positive terms. For this to be small (as demanded by the left-hand side of the equation) one or the other term in the product must be small, for every term in the sum. If $p_{z_\ell}(y,t)$ is small, the implication is that in time $t$ very little probability flows from $z_\ell$ to $y$. However, when $p_{z_\ell}(y,t)$ is not small, all $p$ positive linear forms must be small. This will be satisfied if the individual values $|A_k(z_\ell)-A_k(y)|$ are small. 

So motivated, we define a distance, $d(x,y)$, as the square root of
\numeq{\distanceA}{d_t^2(x,y) \equiv \sum_{k\geq1}
          \frac{\left(A_k(x)-A_k(y)\right)^2}{1-|\lambda_k|^t}  }%
There is both relevant and irrelevant arbitrariness in this definition. For example, use of the square of the $A$s rather than perhaps the fourth power probably does not much affect the grains. Of more significance is the parameter \mquote{t} in the exponent of $\lambda_k$. This will be the time scale for the coarse graining. Through this explicit insertion we acknowledge that although in this article we do propose a well defined, objective coarse graining process, it is nevertheless dependent on a time scale. An alternative (but physically similar) way to insert a time scale would be to truncate the sum in \eq{\distanceA} to (e.g.) $k$ such that $|\lambda_k|^t > 1-\delta$ for some $\delta$.

\subheader{3.2 Transition matrix based measures of dynamical proximity}
In \S3.1 our desire was to use left eigenvectors, which we consider closely related to measurable quantities, to define dynamical proximity. We made use of the fact that left eigenvectors with eigenvalues near 1 take similar values (at \mquote{x} and \mquote{y}) when the quantity $p_{y}(x,t)\equiv (R^t)_{xy}$ was ``large." Now we go directly to $p_{y}(x,t)$.

As before, we have qualitative leeway, both in the functional form and in the choice of the time-scale. We take as a distance the square root of
\numeq{\distanceB}{ D_t^2(x,y) \equiv -\log\left[\max\left(\left(R^t\right)_{xy},\epsilon\right)\right] \;,
\hbox{~~for~} x\neq y \;.  }%
The quantity $\epsilon$ is small and positive and provides a maximum distance, since individual matrix elements of $R^t$ may vanish (for finite $t$). The value of $D_t^2(x,x)$ is irrelevant for our coarse graining considerations and can be taken to be zero. This ``distance" may {\it not\/} be symmetric.

\subheader{3.3 Other definitions}

Other definitions are also possible. For example, let $\widehat R_{xy} \equiv \max\left( R_{xy},R_{yx}\right)$, which is not in general a stochastic matrix. Let $\gamma$ be a path of $n$ steps from $y$ to $x$, without repetitions and $\gamma(\ell)$ its $\ell^{\thsix}$ state. Then one can consider
$$ r_1(x,y) \equiv \max_{\gamma}\prod_{j} \widehat R_{\gamma(j+1) \gamma(j)}
\quad\hbox{or}\quad
r_2(x,y) \equiv \sum_{\gamma} \prod_j \widehat R_{\gamma(j+1) \gamma(j)} $$
Using either of these quantities, a distance can be defined by $\hbox{distance~} \equiv -\log (r)$.

Another measure of distance between states commonly used in studies of stochastic processes is the first hitting time. This measure (or rather its logarithm) would not be additive for tensor products, although it would satisfy an inequality. Like some of our measures it is not necessarily symmetric.

\subheader{3.4 ``Distance" as a metric}

As noted above, what we call ``distance" is not necessarily a metric or distance in the conventional sense [\nr\friedman]. The function \mquote{d} of \eq\distanceA\ clearly is: the triangle inequality follows as it does for any vector product. On the other hand, this metric has the disadvantage of not being additive for tensor products of spaces. By contrast, \mquote{D} of \eq\distanceB\ has good product structure, but is neither guaranteed to be symmetric, nor need it satisfy the triangle inequality. Other definitions mentioned in Sec.\ 3.3 vary in having or not having some of these properties.

\subheader{3.5 Making grains}
As is implicit, we assume that the defining of a distance function provides the basis for selection of coarse grains. One could use known algorithms to choose clusters (which would be the grains) that minimize suitable functions of the internal grain distances [\nr\leonard]. Some cluster algorithms select special points to serve as nuclei. The extremal points $z_\ell$ in our left eigenvector construction (Sec.\ 3.1) certainly present reasonable candidates for this distinction.

For a non-symmetric distance function it would be reasonable to symmetrize by using (e.g.) $\max\left[D_t(x,y),D_t(y,x)\right]$ or $\left[D_t(x,y) + D_t(y,x)\right]$. This process may induce a loss of associativity: coarse graining in two steps can give different grains from those of a single operation (although with a {\it given} hierarchy of grains our coarse graining process of Sec.\ 2.1 {\it is} associative). Additivity of distance for tensor products can also be lost in this symmetrization.

In the present article we both display distances and calculate clusters. Our algorithm is a Monte Carlo scheme, with an annealing protocol, that minimizes the sum of the internal distances within each grain. Generally we specify the number of grains, but allow grain size to vary (as part of the minimization scheme). A coarse graining was accepted when repeated passes with the algorithm did not lower the sum of the internal distances.

\headerandsubheader{4. Brownian motion and diffusion}{4.1 Single particle}
To show the features our definition we first give an example amenable to analytic treatment. Consider Brownian motion (or diffusion) on a ring, with equal probability to step in either direction. The ring consists of $N$ sites, labeled $1, \dots, N$, with periodic boundary conditions. Define the matrix $B$ as
$$ B=\pmatrix{ 0 & 1 & 0 & 0 & 0 \ldots & 0 \cr
               0 & 0 & 1 & 0 & 0 \ldots & 0 \cr
               0 & 0 & 0 & 1 & 0 \ldots & 0 \cr
               \vdots & \vdots & \vdots & &&\vdots \cr
               0 & 0 & 0 & 0 & 0 \ldots & 1 \cr
               1 & 0 & 0 & 0 & 0 \ldots & 0 \cr
}$$
A transition matrix for Brownian motion is $R\equiv \alpha {\bf 1}+({1-\alpha})\left(B+B^{\transpose}\right)/2$, with $0\leq\alpha<1$ and {\sansseriften T} indicating matrix transpose. This matrix is doubly stochastic, satisfies detailed balance, and has equal left and right eigenvectors. The eigenvalues and eigenvectors are
$$\eqalign{ \lambda_k&=\alpha+(1-\alpha)\cos \phi_k \;,
                    \qquad \phi_k=\frac{2\pi k}N\;,\ k=0,1,\dots,\left[\frac N2\right] \cr
u_k(x)&=N^{-1/2}\cos\left(x\phi_k\right) \;, \quad v_k(x)=N^{-1/2}\sin\left(x\phi_k\right) \cr
      & \qquad\qquad k=1,2,\dots,\left[\frac N2\right] \;, \quad x=1,2,\dots,N \cr
}$$
where $[w]$ indicates the integer part of $w$. For even $N$, $v_k$ is zero for $k=0,N/2$, and for odd $N$ it vanishes for $k=0$, in both cases giving the correct number of independent eigenvectors. 

For one dimensional motion of a single Brownian particle there is a simplification. The state space coincides with the configuration space, although in most applications the state space is vastly larger. This is two-edged: it makes our job easier but may obscure a logical distinction. We later discuss more complex spaces built on Brownian motion, and the distinction should then be clear.

We next look at the distance function using the prescription of \eq{\distanceA} with $t=1$. With not too much manipulation one obtains
$$ d^2_1(x,y)=\frac{2}{(1-\alpha)}\frac1N\sum_{k=1}^{[N/2]}
             \frac{\sin^2\left((x-y)\phi_k/2\right)}{\sin^2\left(\phi_k/2\right)} $$
For large $N$ this becomes [\nr{\gradshteyn}]
$$  d^2_1(x,y)=\frac{1}{(1-\alpha)}\frac2\pi\int_0^{\pi/2} \frac{\sin^2((x-y)\theta)}{\sin^2\theta}
              =\frac{|x-y|}{(1-\alpha)} $$
($x$ and $y$ are integers). Thus the coarse grains reconstruct configuration space.

\topinsert 
\def\figsizex{4.0} \def\figsizey{3.26} 
\dimen50=\hsize \advance\dimen50 by -\figsizex truein \divide\dimen50 by 2
\def\innermargin{\dimen50}
\hbox{ 
\lsfig{\figsizex}{\figsizey}{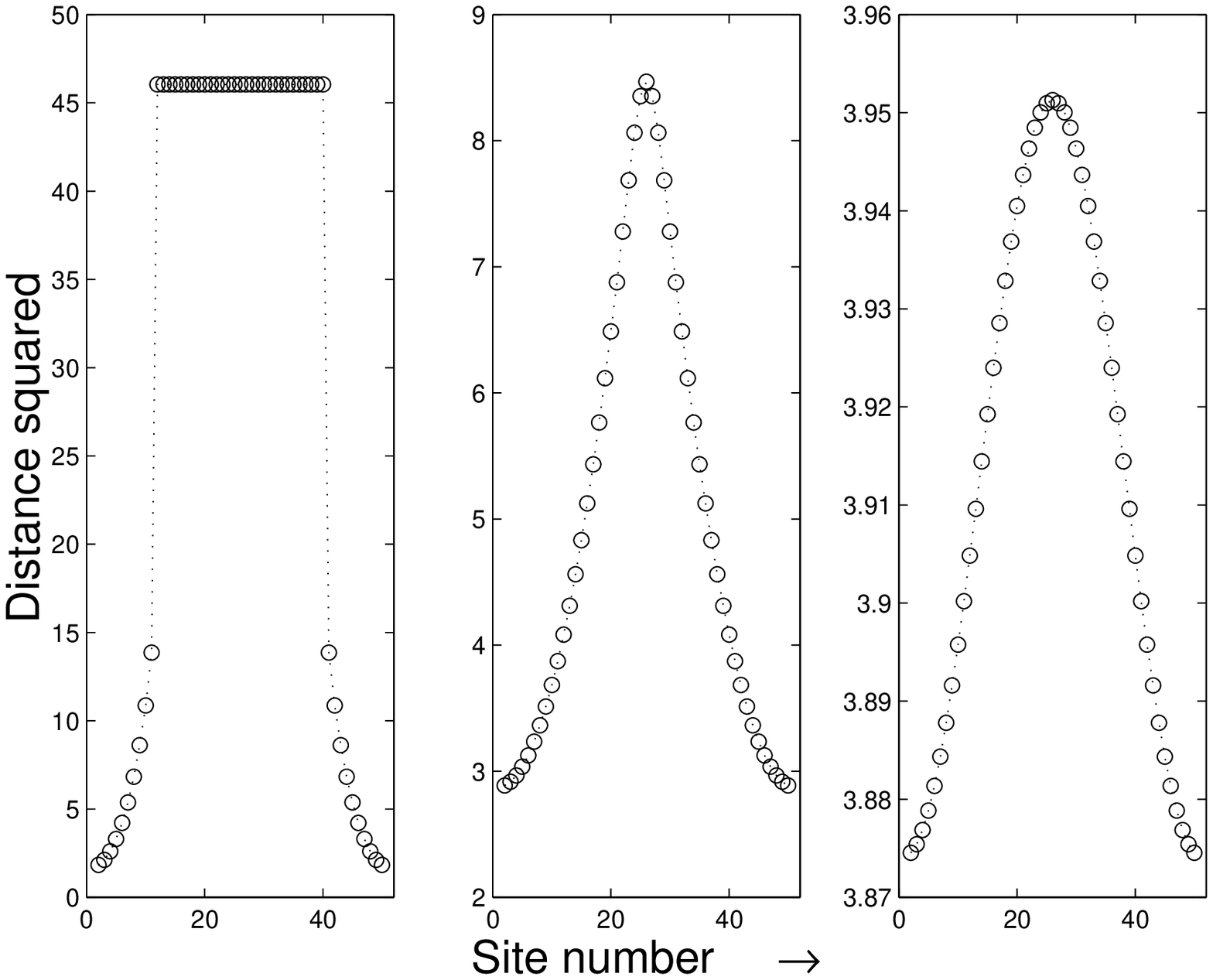}  \hskip \innermargin ~}  
\NI Figure 1. Distance from state \#1 for Brownian motion on 50 sites. The figures differ in their time scale: from left to right, the values for $t$ (of \eq{\distanceB}) are 10, 100, 1000. The maximum distance (cf.\ \eq{\distanceB}) is 46.
\vskip 1 pt
\hbox to \hsize {\vrule  width 6 true in height .4pt depth 0pt}
\endinsert

Our second distance definition can also be approached analytically, bearing in mind that
$$(R^t)_{xy} = \frac1N\sum\left[\alpha +(1-\alpha)\cos\phi_k\right]^t 
                              \sin^2\left(\frac{(x-y)\phi_k}{2}\right)  \;. $$
However, we instead illustrate the results graphically. In Fig.\ 1 we show $R^t_{x1}$ versus $x$ for three different values of the time scale, $t$. Note that all distances are symmetric (recalling our periodic boundary conditions) with respect to site \#1. As before, and in all cases, configuration space is recovered.

The three cases in Fig.\ 1 illustrate three characteristic observational situations. In the leftmost figure there is very good time resolution (relative to the dynamics of the process studied) and sites more than 10 units away from one another will never be put in the same grain by a clustering algorithm that rejects grains containing ``infinitely" separated states. In the middle figure, $t=100$, and although all grains are at ``finite" distance, one can easily distinguish near from far. In the rightmost figure we use $t=1000$, which leads to small value differences in the distance, ranging only from 3.87 to 3.96. In building grains this may be obscured by other factors (if this is part of a larger problem), leading perhaps to the inclusion of all these points in the same coarse grain.

\subheader{4.2 Multiparticle diffusion}

From Brownian motion of a single particle in one dimension, one can extend these results to the diffusion of a large number of particles. The underlying configuration space for all particles and all dimensions is the same as before (the integers, 1 to $N$, with periodic boundary conditions), so that the overall state space is a tensor product of $M$ copies of the configuration space, where $M$ is the number of particles times the number of dimensions.

For this system the coincidental one-to-one correspondence between state-space ($\Omega$) and configuration space no longer holds. We examine the distance \mquote{D} for this case. Microscopic states will be written $x=(x_1,x_2,\dots,x_M)$ and $y=(y_1,y_2,\dots,y_M)$. The transition matrix is $R_{xy} = \prod_\ell R_{x_\ell y_\ell}$. It is immediate from \eq{\distanceB} that
\numeq{\sumofdistances}{D^2_t(x,y)=\sum_\ell D^2_t(x_\ell,y_\ell)  \;,}%
where \mquote{D} is either the 1- or $M$-dimensional distance, depending on its argument. Note that for the definition of \eq{\distanceB} the additive property, ({\sumofdistances}), will be true whenever one builds $\Omega$ as a tensor product.

\midinsert 
\def\figsizex{4.0} \def\figsizey{3.25} 
\dimen50=\hsize \advance\dimen50 by -\figsizex truein \divide\dimen50 by 2
\def\innermargin{\dimen50}   
\hbox{  
\lsfig{\figsizex}{\figsizey}{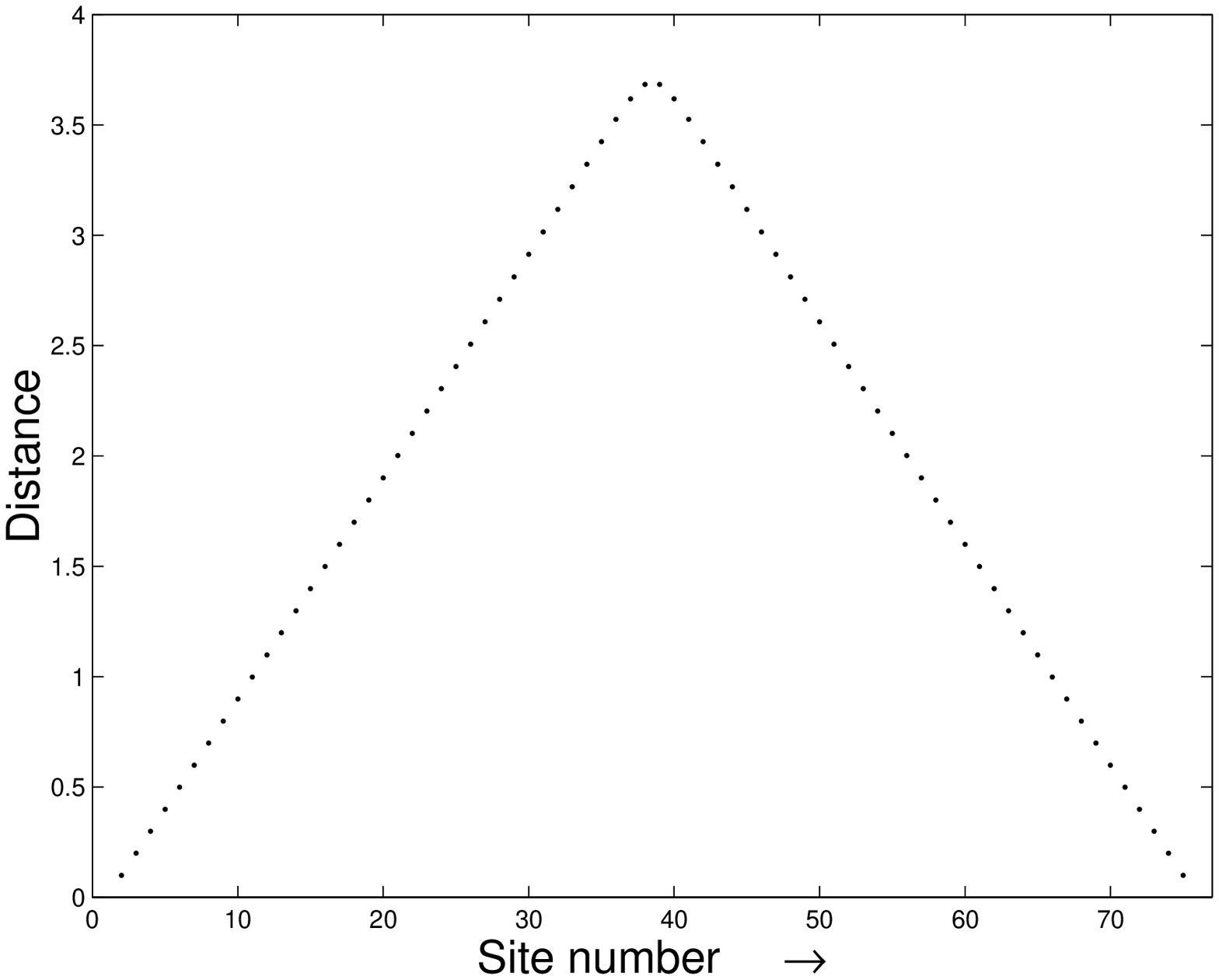} \hskip \innermargin ~}  
\NI Figure 2. Using distances from site \#1 (for $N=75$ and $t=100$, using \eq\distanceB), a constant is subtracted, and the square root taken. The result is plotted and is evidently well approximated by a straight line near the point, ``1."
\vskip 1 pt
\hbox to \hsize {\vrule  width 6 true in height .4pt depth 0pt}
\endinsert

As a result of \eq\sumofdistances, coarse grains in $\Omega$ can be taken to be compact objects (perhaps hypercubes) in {\bf R}$^M$. The function \mquote{D} is in fact close to being quadratic in distance (for points not too far apart). See Fig.~2, where, after subtracting an appropriate constant, we take the square root of this distance, a process that evidently gives a good approximation to a straight line.

Using a coarse graining depending on this \mquote{D} alone does not lead to a notion of density, since the particles are distinguishable. This means that so long as each pair $x_\ell$ and $y_\ell$ are close, the corresponding $x$ and $y$ will be close---despite the fact that the various values of $x_\ell$ (for different $\ell$) are {\it not\/} close. Thus the ``center of the swarm" or other such collective variables do not emerge from this distance function. This kind of emergence will be seen in other cases, as described below.

\header{5. Dynamic Ferromagnet}

The one-dimensional Ising model has the energy function
\numeq{\ising}{H=-J\sum_{k=1}^N \sigma_k\sigma_{k+1} -B\sum_{k=1}^N \sigma_k \;,}%
where $\sigma_k$ is a $\pm1$-valued spin at site \#$k$, $J>0$ is the spin-spin coupling constant, and $B$ the external magnetic field. Index addition in \eq\ising\ is modulo~$N$. Many stochastic-dynamical schemes are consistent with this energy and we use the following. A site, $k'$, is randomly selected and the change in energy that would occur {\it if\/} that spin were flipped, $\Delta E\equiv H(\sigma_{k'}\to-\sigma_{k'}) - H(\hbox{original})$, is calculated. The flip is then implemented with probability $\alpha \exp(-\Delta E/2T)$, where $T$ is a temperature parameter and $\alpha$ a global constant chosen small enough so that all components of the transition matrix are non-negative. A second allowed process is a double flip by opposite pointing neighbor spins. The probability that this occurs is governed by an additional controllable parameter. In this model there is no breakdown in analyticity, but for $T\leq1$ and $B=0$ the system will spend most of the time strongly polarized. (The stationary, equilibrium distribution as a function of magnetization is strongly peaked at symmetric nonzero values.) Above $T=2$ the system is found mostly at small magnetization values.

\midinsert 
\def\figsizex{2.5} \def\figsizey{2.055} 
\def\moveleftstart{-.1truein} 
\hbox{  \hskip \moveleftstart
\lsfig{\figsizex}{\figsizey}{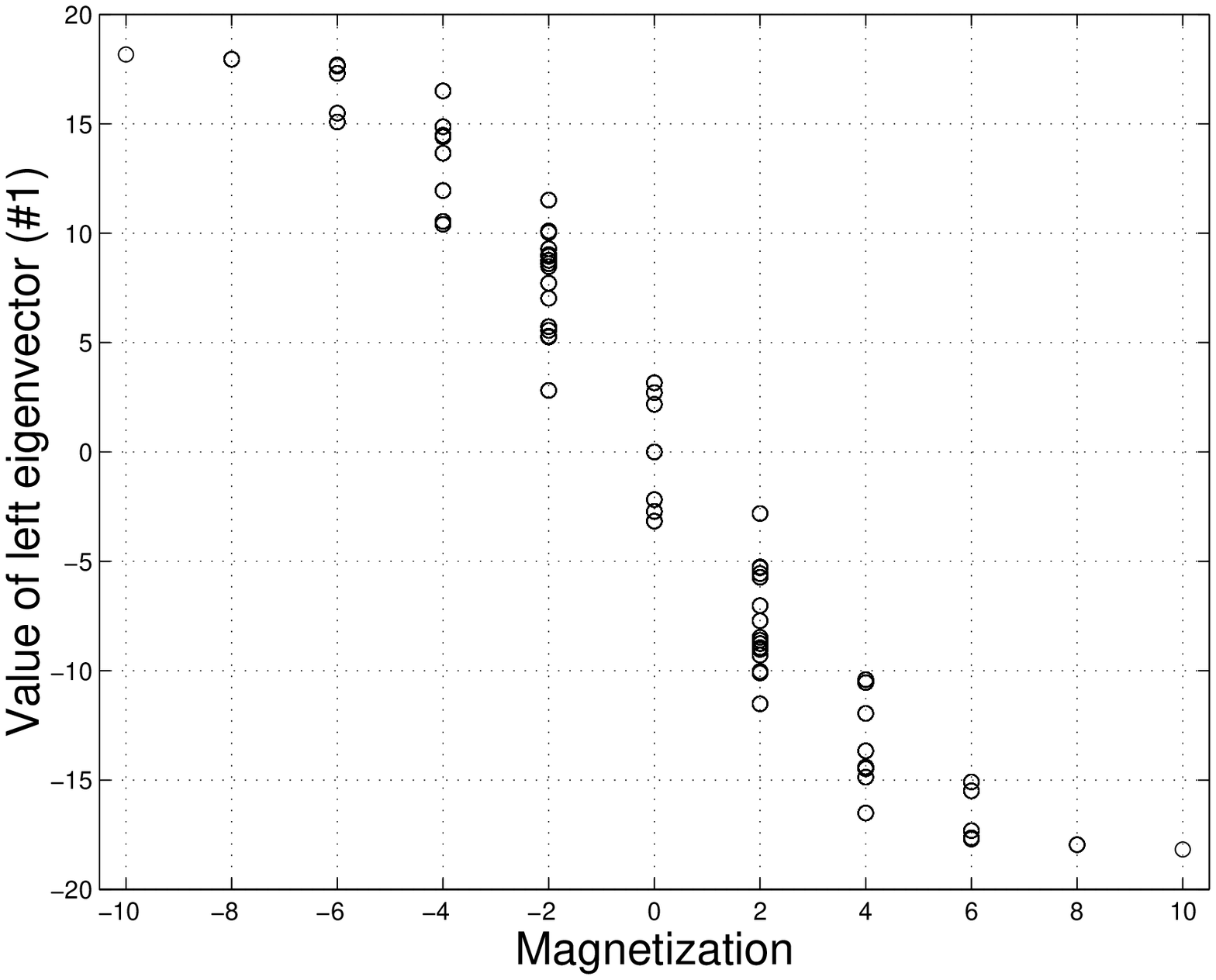}
\lsfig{\figsizex}{\figsizey}{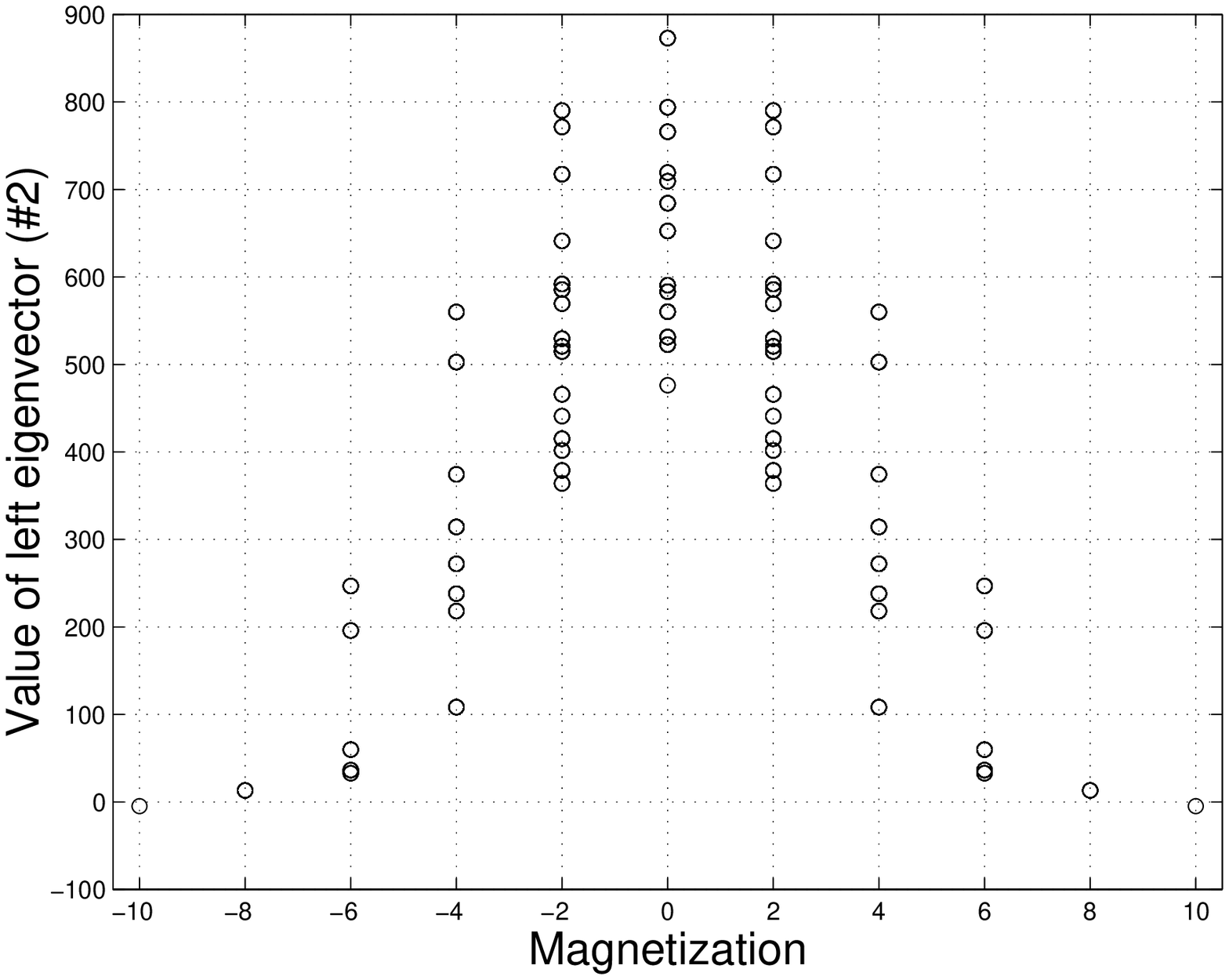}  }
\NI Figure 3. Values of the left eigenvectors, $A_1$ and $A_2$, as a function of magnetization. Temperature is 0.6. Ten spins are used and $B=0$.
\vskip 1 pt
\hbox to \hsize {\vrule  width 6 true in height .4pt depth 0pt}
\endinsert

\midinsert 
\def\figsizex{4.0} \def\figsizey{3.2} 
\dimen50=\hsize \advance\dimen50 by -\figsizex truein \divide\dimen50 by 2
\def\innermargin{\dimen50}
\hbox{  
\lsfig{\figsizex}{\figsizey}{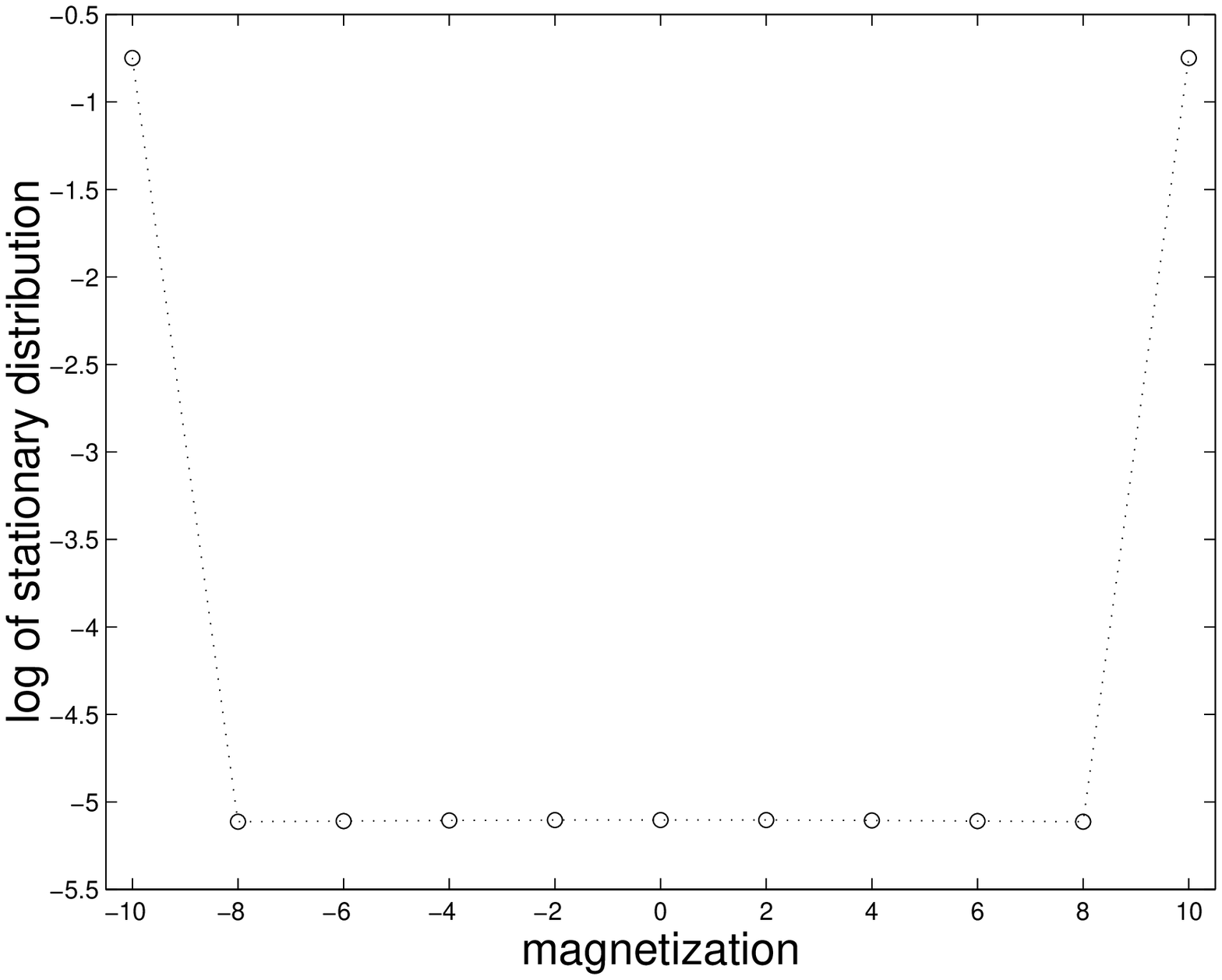} \hskip \innermargin ~}  
\NI Figure 4. Probability distribution as a function of magnetization for the one-dimensional Ising model at $T=0.6$ and $B=0$.
\vskip 1 pt
\hbox to \hsize {\vrule  width 6 true in height .4pt depth 0pt}
\endinsert

The transition matrix, $R$, in this case is $2^N\!\times2^N$, but it is sparse, so it is not a problem to look at $N=10$ or larger on a \hbox{PC}. As expected, for $T\leq1$ there is near degeneracy of $1\equiv\lambda_0$ and $\lambda_1$. (As $T$ increases above 1, the gap, $1-\lambda_1$, rapidly increases.)

In Fig.\ 3 we plot the values of the left eigenfunctions, $A_1$ (whose eigenvalue is nearly degenerate with 1 for low temperature) and $A_2$, as a function of magnetization. That is, while there are $2^N$ states, the magnetization ($\sum \sigma_k$) only takes $N+1$ values, and it is these that we place on the abscissa. At this temperature, the first few gaps ($1-\lambda$) are 1.3$\times 10^{-5}$, 4.0$\times 10^{-3}$, 6.0$\times 10^{-3}$ (twice), 1.2$\times 10^{-2}$ (twice), 1.3$\times 10^{-2}$. Thus the distance function \mquote{d} will be heavily influenced by $A_1$. It is clear from the figure that magnetization is an emergent macroscopic variable. Not only are the two extreme magnetizations sharply distinguished, but for differences of size 4 there is no overlap in $A_1$ values. If this is combined with information from $A_2$ (on the right in Fig.\ 3), an additional separation can be made between large magnetization (of either sign) and small magnetization states. The stationary state is in one of two highly polarized states, and while conventional definitions do not attribute to this system a transition, by criteria more suitable for finite systems (in particular those of [\firstor]) there is such a transition. For reference, in Fig.\ 4 we plot the equilibrium probability as a function of magnetization.

\topinsert 
\def\figsizex{2.5} \def\figsizey{2.466} 
\def\moveleftstart{-.1truein}
\hbox{  \hskip \moveleftstart
\lsfig{\figsizex}{\figsizey}{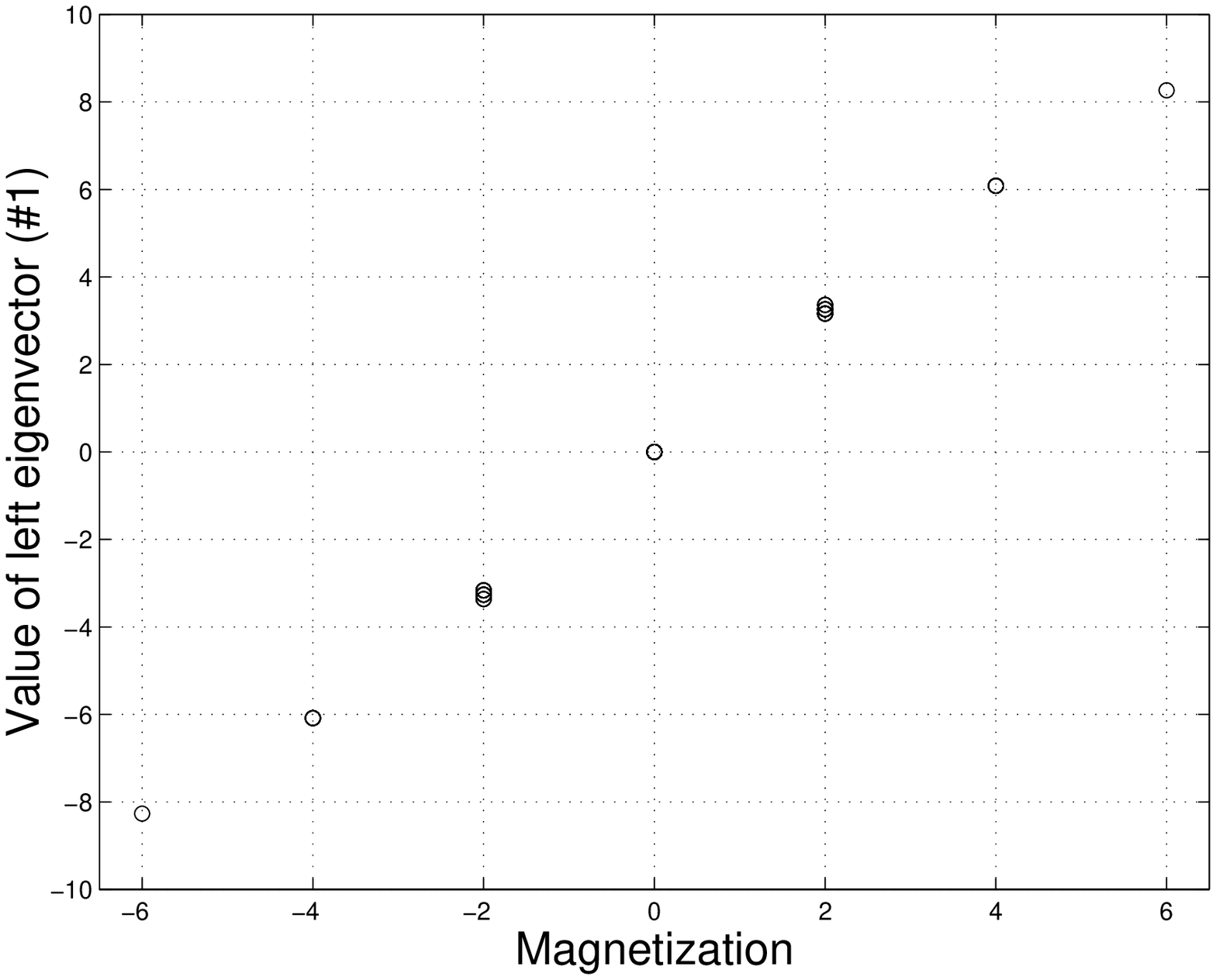} 
\lsfig{\figsizex}{\figsizey}{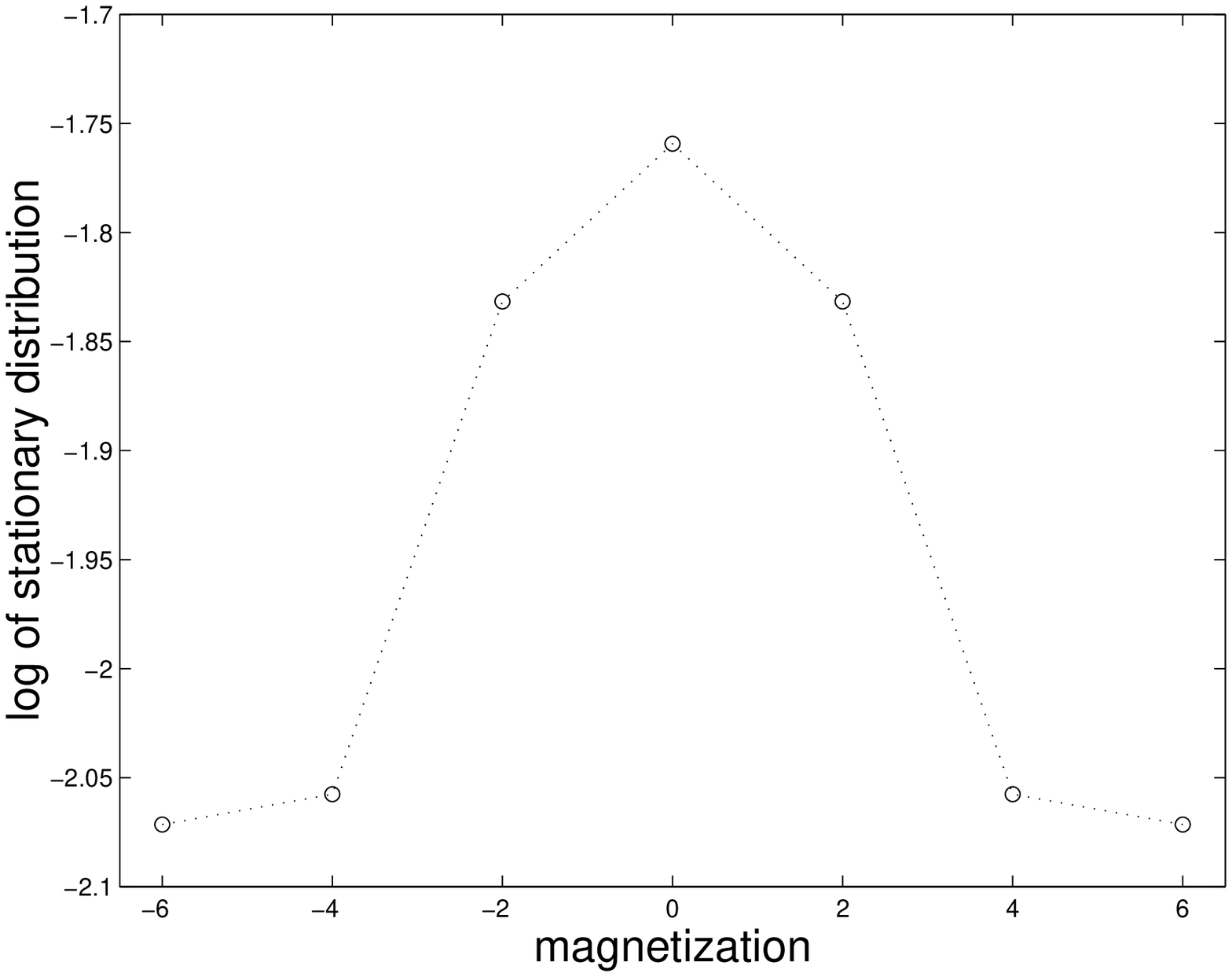}
}  
\NI Figure 5. Value of the left eigenvector, $A_1$, and probability distribution of the stationary state, as a function of magnetization. Temperature is 2.25. Six spins are used and $B=0$.
\vskip 1 pt
\hbox to \hsize {\vrule  width 6 true in height .4pt depth 0pt}
\endinsert

Even for temperatures such that the system is not highly polarized and is mostly found at small values of magnetization, one still finds that magnetization is an emergent macroscopic variable. This can be seen in Fig.\ 5, where the eigenfunction $A_1$ is shown as well as the probability distribution, both as functions of magnetization. The distinction in $A_1$ values among different magnetization states is even stronger than at low temperature, but in practice this eigenvector will not weigh so heavily in the distance function because the lead eigenvector is far from degenerate. Specifically the first few gaps ($1-\lambda$) are now 0.015, 0.051, 0.10, 0.15 (twice), 0.16, 0.19. This is in sharp contrast to the low temperature situation where the smallest gap was orders of magnitude smaller than the next one.

\topinsert 
 \def\figsizex{0.9} \def\figsizey{4}
 
 \def\innermargin{.5truein}
 \hbox{ \hskip 1.5 truein
 \lsfig{\figsizex}{\figsizey}{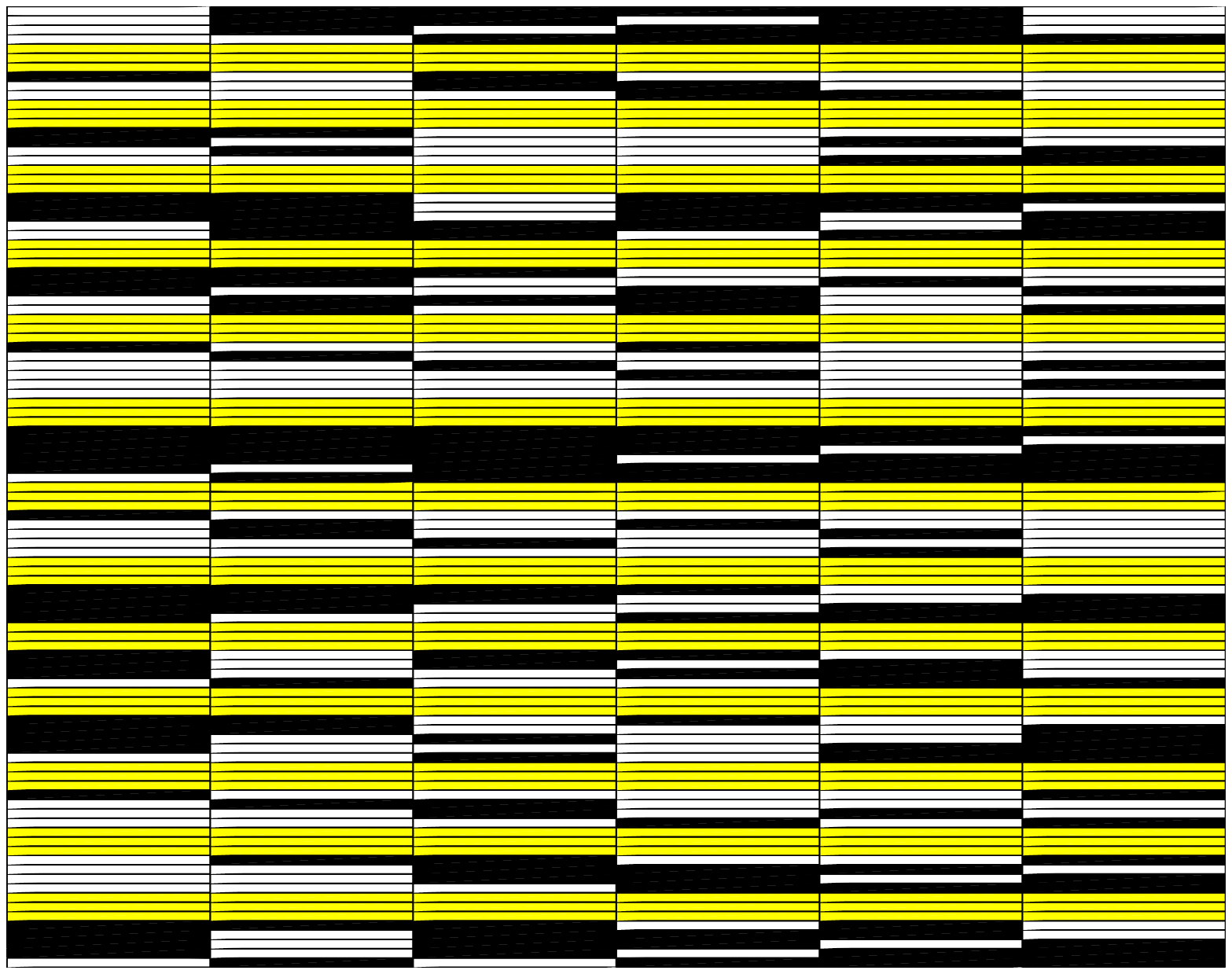} \hskip \innermargin
 \lsfig{\figsizex}{\figsizey}{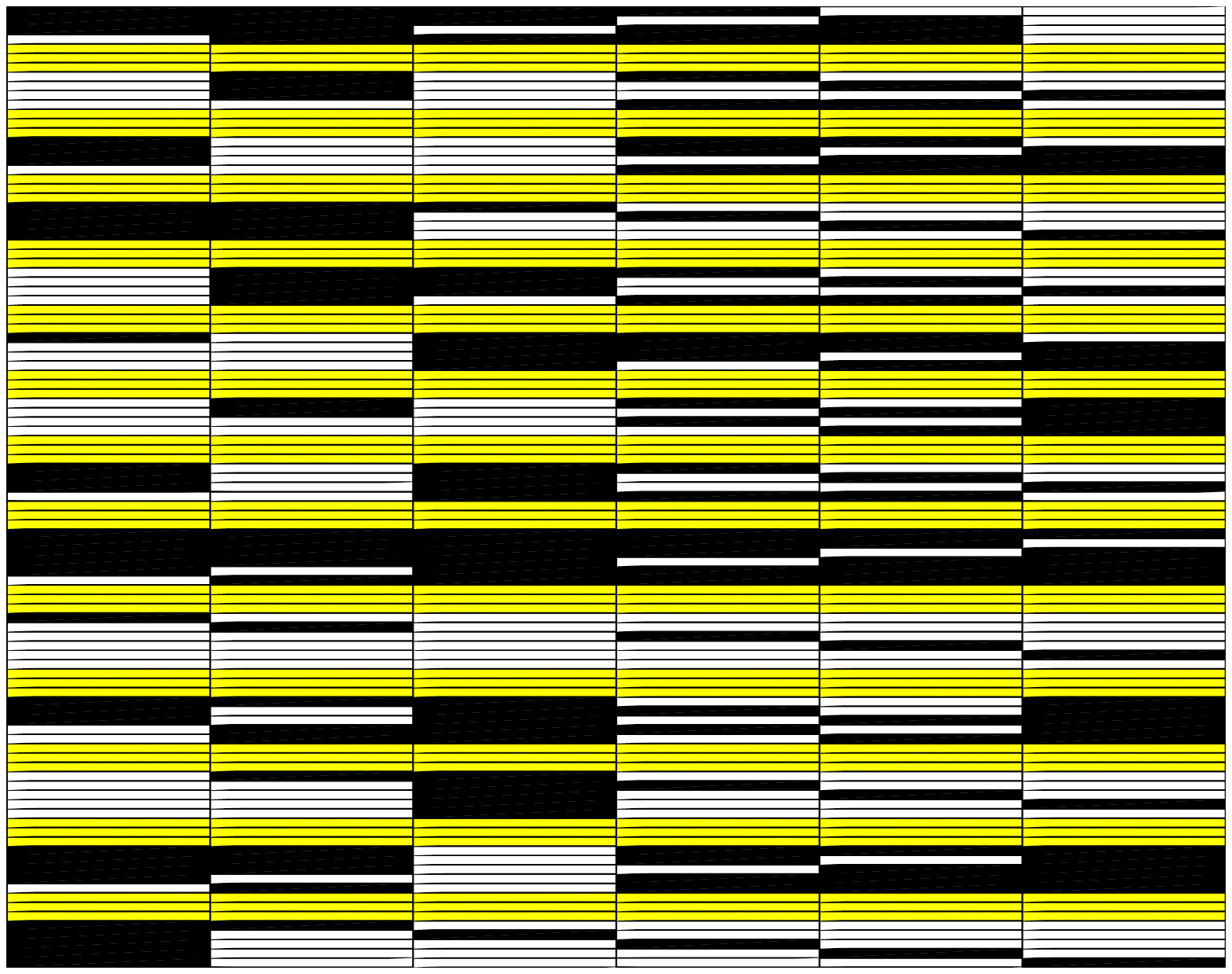}
 }  
 \NI Figure 6. Coarse grains, as computed from the distance function given in \eq{\distanceA}, using a cluster internal distance minimization algorithm. The system is the one dimensional Ising model at temperatures 0.8 (left) and 2.25. The grains are read from the figures in the following way. Each row represents a state of the system, with blackened regions corresponding to excited sites. Three empty grayish rows separate each pair of grains.

\vskip 1 pt
 \hbox to \hsize {\vrule  width 6 true in height .4pt depth 0pt}
 \endinsert

For this example we went a step further and calculated the actual grains. The distance (squared) matrix derived from \eq{\distanceA} was fed into an algorithm designed to minimize the sum of internal cluster distances. What emerged was a sorting according to magnetization, both above and below the transition. In Fig.\ 6 we show the grains that result from this procedure. The magnetization of the states {\it within} each grain is given in Table~1.

\midinsert
\begingroup
\settabs 2 \columns
\+ High temperature ($T=2.25$)  & Low Temperature ($T=0.8$)\cr
\settabs 14 \columns
\def\ph{\phantom{-}}
\+ \ph2& \ph2& \ph2& \ph2&&     &&   \ph2&\ph0&\ph0&\ph2& \cr
\+ -2&-2&-2&-2&           &     &&   -2&-2&-2&  \cr
\+ \ph0&\ph0&\ph0&\ph0&&             &&-2&-2&-2&-2& \cr
\+ \ph0&\ph0&\ph0&\ph0&&             &&\ph4&\ph2&\ph2&\ph2&\ph2& \cr
\+ \ph0&\ph0&\ph0&\ph0&&             &&\ph0&\ph0&\ph0&\ph0&\ph0&  \cr
\+ \ph2&\ph2&\ph0&\ph0&&             && -2&-4&-2&-4&-4&-6& \cr
\+ \ph0&\ph0&-2&-2&&           &&\ph6&\ph4&\ph4&\ph4&\ph4&\ph4& \cr
\+ \ph0&\ph0&\ph0&\ph0&&             &&-4&-2&-2&-4&-4& \cr
\+ \ph6&\ph4&\ph4&\ph4&\ph4&\ph4&         &\ph2&\ph2&\ph2&\ph2& \cr
\+ -4&-4&-4&-4&-4&-6&   &\ph0&\ph0&\ph0&\ph0& \cr
\+ \ph2&\ph2&\ph2&\ph2&\ph2&           && \ph0&\ph0&\ph0&\ph0&\ph0& \cr
\+ -2&-2&-2&-2&-4&      &&-2&-2&-2&-2& \cr
\+ \ph4&\ph2&\ph2&\ph2&\ph2&           &&\ph 0&\ph0&\ph0&\ph0& \cr
\+ -2&-2&-2&-2&-2&      &&\ph2&\ph2&\ph2&\ph2&\ph2 \cr
\endgroup
\NI Table 1. Magnetization of the states within a grain. Each row of the left and right tables corresponds to a grain and lists the magnetization of each state in that grain. Note that within each grain the values of magnetization are almost all equal, demonstrating that with the definition (\distanceB) the order parameter, magnetization, is an emergent quantity.
\vskip 1 pt
 \hbox to \hsize {\vrule  width 6 true in height .4pt depth 0pt}
\endinsert

\header{6. Heat flow}

Heat flow represents the diffusion of phonons and we model this in the following way. $N$ 2-level systems lie on a line. The two levels at each site will be called up and down, excited or unexcited. At one end of the line is a source of heat. The stochastic rule is that if the left-hand site is down, there is some probability that it will become up, absorbing energy from the hot reservoir. That probability is related to the temperature of the source. On the right the opposite can happen. If the rightmost site is up, there is some probability that it will drop, modeling the passage of that energy to the cold reservoir. At all other sites there is no preferred direction. An adjacent pair of sites is randomly selected and if one is up and the other down, there is a fixed probability that they will exchange levels of excitation. With this rule there is linear temperature dependence from one end to the other (where ``temperature" means time-averaged excitation).

 \topinsert 
 \def\figsizex{0.9} \def\figsizey{4}
 
 \def\innermargin{.5truein}
 \hbox{
 \lsfig{\figsizex}{\figsizey}{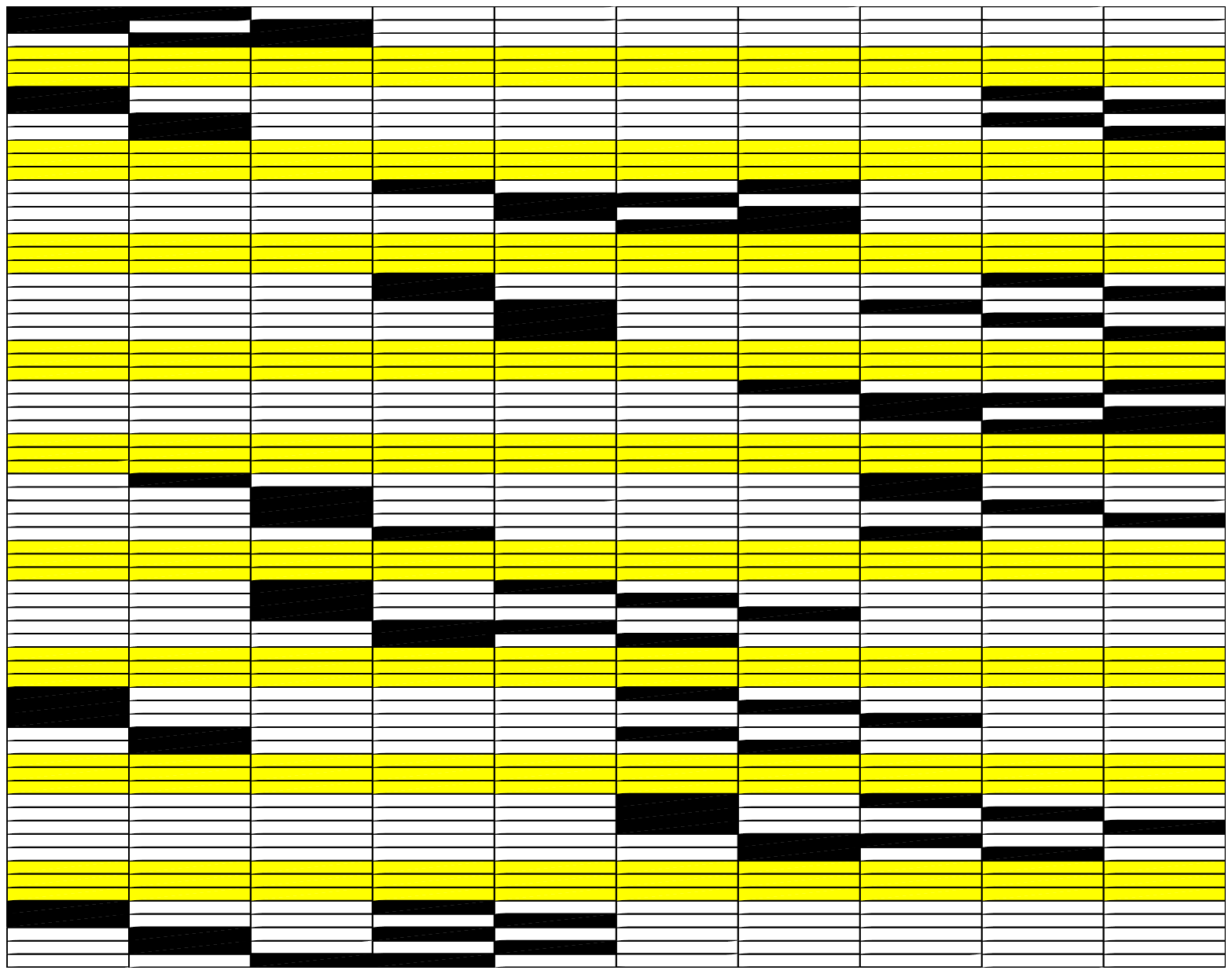} \hskip \innermargin
 \lsfig{\figsizex}{\figsizey}{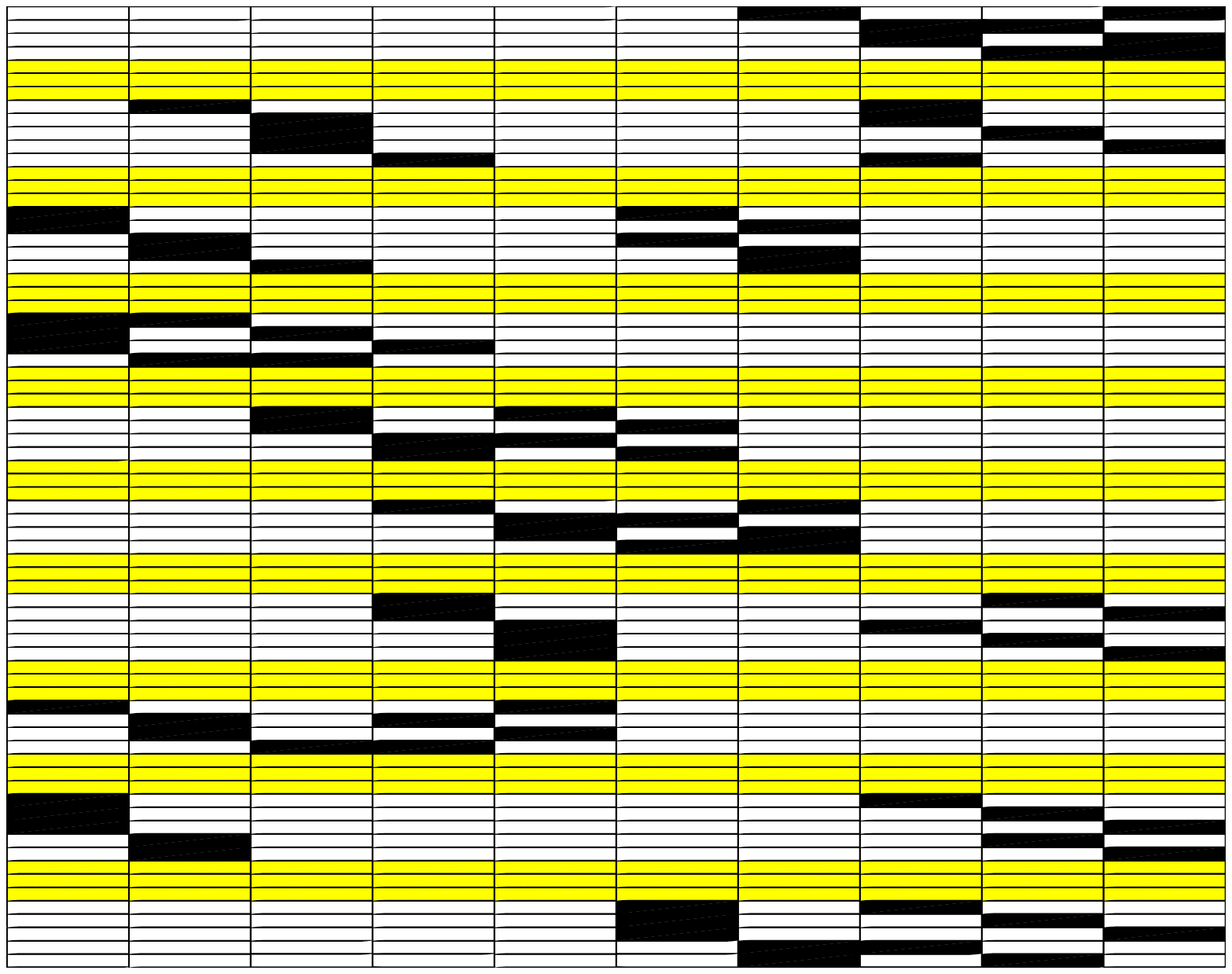} \hskip \innermargin
 \lsfig{\figsizex}{\figsizey}{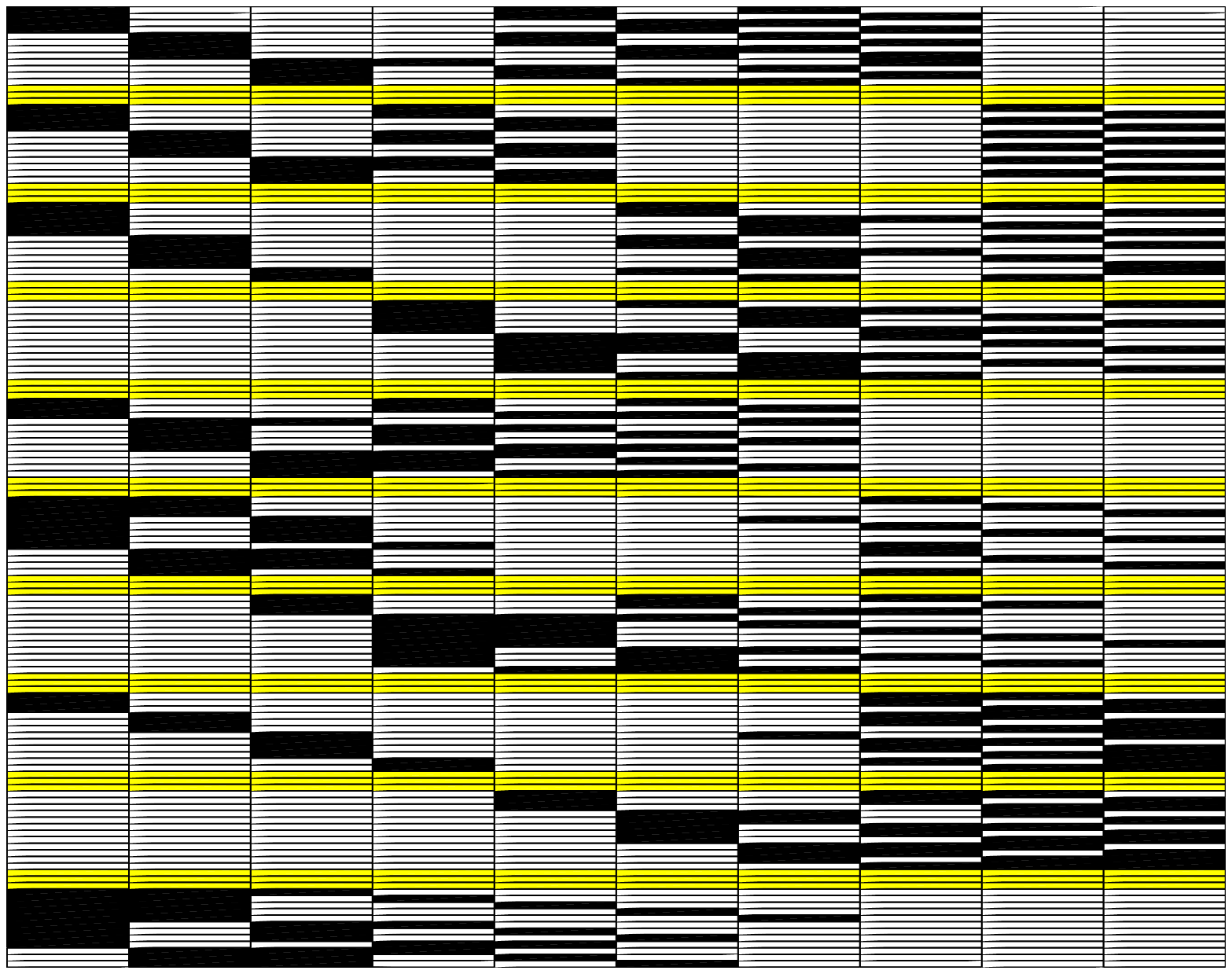} \hskip \innermargin 
 \lsfig{\figsizex}{\figsizey}{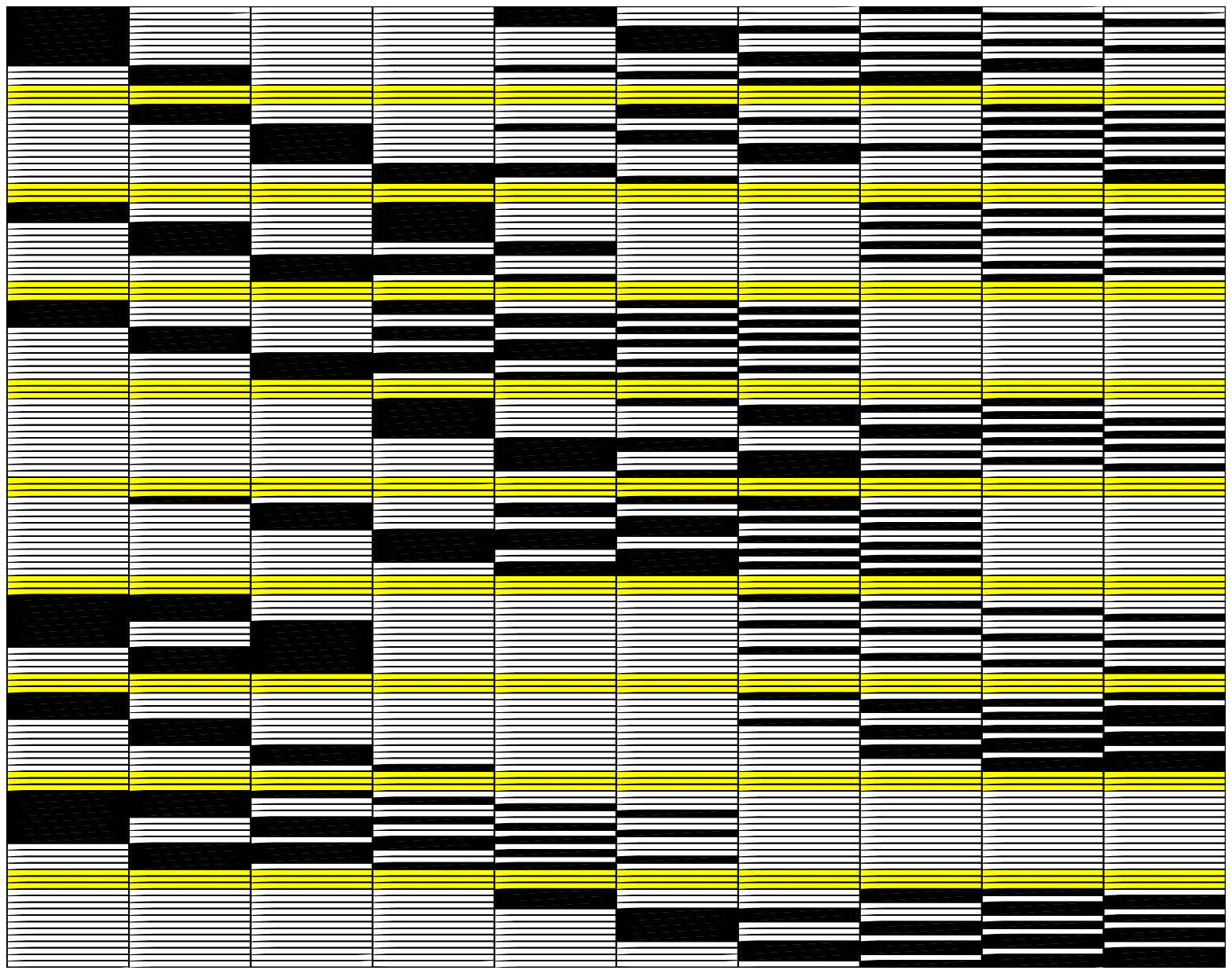} 
 }  
 \NI Figure 7. Coarse grains, as computed from the distance function given in \eq{\distanceA}, using an internal cluster distance minimization algorithm. Two parameters are varied: $t$, the time scale, and $M$ the number of sites excited. The four figures correspond respectively to $(t,M)=(1,2), (10,2), (1,3), (10,3)$. The grains are read from the graphs as in the previous figure. Thus for $t=1$, $M=2$ the first grain corresponds to the upper 3 rows of the leftmost figure. All states have two excited sites and in this grain, 2 out of the first 3 sites are excited in all cases.
\vskip 1 pt
 \hbox to \hsize {\vrule  width 6 true in height .4pt depth 0pt}
 \endinsert

Coarse graining of the entire state space tends to put states of the same total excitation into the same grains. We provide data here on a more detailed kind of coarse graining, related to density. Restrict attention to subspaces with a fixed level of excitation. Among these we define a distance using \eq{\distanceA}. Not unexpectedly, grains tend to contain states with similar {\it spatial\/} distribution of excitation. In Fig.\ 7 we show the grains for the case of 10 sites and excitation levels 2 (for which there are 45 states) and 3 (with 120). Once again coordinate space emerges as a relevant macroscopic quantity. There is also the beginning of the emergence of a macroscopic ``density."

\header{7. Discussion}

Coarse graining based on dynamics has been shown to recover at least some aspects of the macroscopic world, in particular it accounts for the emergence of some of the less obvious collective variables. The ``obvious" variables correspond to configuration space and indeed they do emerge in a natural way. But others are well defined by virtue of collective behavior, a feature that dynamical criteria can select. We have shown that density and magnetization fall into this category.

But the story is far from complete. We did not argue for a unique choice of distance function on the global state space, but rather offered several possibilities. Our criteria for choosing were mainly theoretical and aesthetic, additivity over tensor products or symmetry. Ideally these differences would be expressed in the laboratory, for example by changing definitions of coarse grains, one changes entropy and ultimately temperature, through $\frac1T=\frac{\partial S}{\partial E}$. However, the enormous numbers involved in entropy definitions---which has permitted rather loose definitions of coarse grains---will probably preclude easy experimental tests. A second kind of ambiguity relates to the selection of a time scale. Here too one would hope for physical input, although in this case one can easily imagine that as the time scale changes, what one (coarser) observer would call a fluctuation another would consider deterministic time evolution.

This last remark emphasizes that the quantitative implementation of coarse graining that we have initiated in this paper moves the Second Law of Thermodynamics from a law of Nature to a convenient tautology [\nr\timebook]. Another perspective on this can be seen by recalling an observation in [\effect]. In that article the following is shown: {\it Coarse graining a distribution function, evolving it forward, and then again coarse graining, either increases the entropy or leaves it unchanged.} This is effectively a ``proof" of irreversibility, but the demonstration relied on an important physical assumption, to wit, by the time the second coarse graining is implemented, {\it there must be local equilibrium within each grain}. Of course this is the same point that is made in [\landau]. But the implication is that with faster observations there will no (or less) entropy increase [\nr\details].

In the introductory section we gave as an example of coarse grain ambiguity the ability or inability, as the case may be, to distinguish $^{12}$C from $^{14}$\hbox{C}. How is that addressed by our ideas? Consideration of this example forces the realization that the ``dynamics" that enters our matrix \mquote{R} must include the observer. An observer that is able to distinguish these isotopes will do so by inducing microscopic interactions that percolate upward to the macroscopic level. In a way this is like the situation in quantum mechanics, but with no a priori limitation on the weakness of the interaction.

\header{Acknowledgements}
This article is dedicated to Martin Gutzwiller on the occasion of his 75th birthday. We are grateful to L. J. Schulman for useful discussions. The research was supported in part by the United States National Science Foundation grant PHY 97 21459.

\header{References}

\pritem{\kac} M. Kac, {\it Probability and Related Topics in Physical Sciences},   Interscience, London (1959). Proceedings of the Summer Seminar, Boulder, Colorado, 1957, with  special lectures by G. E. Uhlenbeck, A. R. Hibbs and B. van der Pol.

\pritem{\mackey} For additional objections to coarse graining see M. C. Mackey, Microscopic Dynamics and the Second Law of Thermodynamics, in {\it Proceedings of the conference, ``Time's Arrows, Quantum Measurements and Superluminal Behavior," Naples, Italy, October 2000}, D. Mugnai, A. Ranfagni and L. S. Schulman, eds., Italian National Research Council (CNR). See also M. C. Mackey, {\it Time's Arrow: The Origins of Thermodynamic Behavior}, Springer, New York (1992).

\pritem{\zeh} H. D. Zeh, {\it The Physical Basis of The Direction of Time}, 3rd ed., Springer, New York (1999).

\pritem{\hartle} J. B. Hartle, Spacetime coarse grainings in nonrelativistic quantum  mechanics, Phys.\ Rev.\ D {\bf 44}, 3173 (1991).

\pritem{\halliwell} J. J. Halliwell, The Emergence of Hydrodynamic Equations from Quantum Theory: A Decoherent Histories Analysis, Found.\ Phys.\ Lett.\ {\bf 39}, 1767 (2000).

\pritem{\effect} L. S. Schulman, Causality is an effect, in {\it Proceedings of the conference, ``Time's Arrows, Quantum Measurements and Superluminal Behavior," Naples, Italy, October 2000}, D. Mugnai, A. Ranfagni and L. S. Schulman, eds., Italian National Research Council (CNR).

\pritem{\master} B. Gaveau and L. S. Schulman, Master equation formulation of non-equilibrium statistical mechanics, J. Math.\ Phys.\ {\bf 37}, 3897 (1996).

\pritem{\framework} B. Gaveau and L. S. Schulman, A general framework for non-equilibrium phenomena: The master equation and its formal consequences, Phys.\ Lett.\ A {\bf 229}, 347 (1997).

\pritem{\firstor} B. Gaveau and L. S. Schulman, Theory of non-equilibrium first order phase transitions for stochastic dynamics, J. Math.\ Phys.\ {\bf 39}, 1517 (1998).

\pritem{\signat} B. Gaveau, A. Lesne and L. S. Schulman, Spectral signatures of hierarchical relaxation, Phys.\ Lett.\ A {\bf 258}, 222 (1999).

\pritem{\landau} L. D. Landau and E. M. Lifshitz, (1980) {\it Statistical Physics}, 3rd ed., Part 1, pp.\ 13--14. Translated by J. B. Sykes and M. J. Kearsley, Course of Theoretical Physics, Volume 5, Pergamon Press, Oxford (1980).

\pritem{\currentdef} The current matrix for a given $R$ (and its $\pzero$), is $J_{xy} \equiv R_{xy} \pzero(y) - R_{yx} \pzero(x)$.

\pritem{\noeigs} Of course clustering may not occur. In its absence our coarse graining scheme breaks down.

\pritem{\proveindep} Here is a proof that \hullA\ is of codimension 0 in \hbox{\Rp}. Suppose otherwise. Then all $V\in\cal A$ would lie on a hyperplane. Let this hyperplane be defined by constants $\nu_\ell$, $\ell=1,\dots,p$ (and not all zero), so that for $\xi$ in the hyperplane, $\sum \nu_\ell \xi_\ell =0$ (the constant that ordinarily appears on the right, defining the hyperplane, is 0 because 0 is in \hullA). It follows that for every $x\in\Omega$, $\sum_\ell \nu_\ell A_\ell(x)=0$. This would imply linear dependence of the $A$s in the $N$-dimensional space that $R$ acts on.

\pritem{\friedman} A. Friedman, {\it Foundations of Modern Analysis}, Dover, New York
(1982).

\pritem{\leonard} L. J. Schulman, Clustering for Edge-Cost Minimization, in {\it Proceedings of the 32nd STOC}, 2000.

\pritem{\gradshteyn} I. S. Gradshteyn, I. M. Ryzhik, {\it Table of Integrals, Series, and Products}, 4th ed., Academic Press, New York (1980).

\pritem{\timebook} L. S. Schulman, {\it Time's Arrows and Quantum Measurement}, Cambridge University Press, Cambridge (1997).

\pritem{\details} One can carry this perspective to its logical extreme, perhaps leading to a physics analog of the remark of L. Mies van der Rohe, ``God is in the details" (New York Herald Tribune, 28 June 1959).

\end

%% file: macros.tex
\font\romsix=cmr6 scaled\magstep0 
\font\sansserifeight=cmss8 scaled\magstep0
\font\romnine=cmr9 scaled\magstep0 
\font\italnine=cmti9 scaled\magstep0 
\font\ttnine=cmtt9 scaled \magstep0
\font\sansseriften=cmss10 scaled\magstep0
\font\romtwelve=cmr10 scaled\magstep1
\def\ident#1{}
\def\frac#1#2{{#1 \over #2}}
\def\header#1{{
  \removelastskip\vskip 20pt plus 40pt \penalty-200 \vskip 0pt plus -32pt
  \NI\bf #1}\nobreak\medskip\nobreak}
\def\subheader#1{\removelastskip
  \vskip 10pt plus 20pt \penalty-200 \vskip 0pt plus -16pt
             \NI{\it #1}\nobreak\smallskip}
\def\headerandsubheader#1#2{{
  \removelastskip\vskip 20pt plus 40pt \penalty-200 \vskip 0pt plus -32pt
  \NI\bf #1}\nobreak\medskip\nobreak\NI{\it #2}\nobreak\smallskip}
\def\standardpage{\hsize=6 true in \vsize= 8.5 true in \hoffset=0.25 true in}
\def\tenstart{\lineskiplimit=1pt\normalbase \tenrm ~~~~~~\par} 
\overfullrule=0pt
\def\negonemu{\mskip -1mu}
\def\eq#1{Eq.\ (#1)} 
\def\ellip{$\ldots$}
\def\NI{\noindent}
\def\ordinalth{ ^{\hbox{\romsix th}} }
\def\pritem#1{\vskip .04in\item{[#1{\ttnine\ident#1}]}}

\newcount\noteno
\def\numfoot#1{\advance \noteno by 1 \footnote
   {$^{\the\noteno}$} {\scrunch #1\toe}}

\newcount\ftnum  \newcount\checknum  \newif\ifeqnerr
\def\aq#1{\global\advance\ftnum by 1 \xdef#1{\the\ftnum}}
\def\neweq#1{{\global\advance\checknum by 1
   \edef\chqtmp{\the\checknum}
   \checkeq#1}}
\def\checkeq#1{\ifx#1\chqtmp\global\eqnerrfalse
   \else \global\eqnerrtrue
        \message{ALLOCATION ERROR for \string#1:
             preassigned #1, in sequence \chqtmp.}\fi}

\def\aqx#1#2{\xdef#2{\the\ftnum#1}}
\def\neweqx#1#2{{\edef\chqtmp{\the\checknum#1}
    \checkeq#2}}

\newcount\eftnum  \newcount\echecknum  \newif\ifeeqnerr
\def\eaq#1{\global\advance\eftnum by 1 \xdef#1{\the\eftnum}}  
\def\neweeq#1{{\global\advance\echecknum by 1
   \edef\chqetmp{\the\echecknum}
   \checkeeq#1}}
\def\checkeeq#1{\ifx#1\chqetmp\global\eqnerrfalse
    \else \global\eqnerrtrue
        \message{ALLOCATION ERROR for \string#1:
             preassigned #1, in sequence \chqetmp.}\fi}

\def\eaqx#1#2{\xdef#2{\the\eftnum#1}}
\def\neweeqx#1#2{{\edef\chqetmp{\the\echecknum#1}
    \checkeeq#2}}

\def\nr#1{\aq#1#1{\ttnine\ident#1}}
\def\numeq#1#2{\eaq#1 {\ttnine\ident#1} $$#2 \eqno(#1)$$ }
\def\mquote#1{``$#1$"}

\input epsf.sty

\def\lsfig#1#2#3{\epsfxsize=#1in \epsfysize=#2in \epsffile{#3.eps}}
 
\def\ptilde{{\tilde p}_0}
\def\pzero{p_{0}}
\def\xtilde{\tilde x} \def\ytilde{\tilde y}
\def\omtilde{\widetilde \Omega} \def\rtilde{\widetilde R}
\def\transpose{\hbox{\sansserifeight T}}
\def\thsix{\hbox{\romsix th}}
\def\hullA{$\widehat{\cal A}$}
\def\Rp{{\bf R}$^p$}